\documentclass[12pt]{iopart}

\usepackage{color}
\usepackage{graphicx}% Include figure files
\usepackage{bm}% bold math

\begin{document}

\title[Distributional Behavior of Diffusion Coefficients]{Distributional Behavior of Diffusion Coefficients Obtained by Single Trajectories 
 in Annealed Transit Time Model}

\author{Takuma Akimoto \& Eiji Yamamoto}

\address{Graduate School of Science and Technology, Keio University, Yokohama, 223-8522, Japan}
\ead{akimoto@keio.jp}
\vspace{10pt}
\begin{indented}
\item[]February 2014
\end{indented}

\begin{abstract}
Local diffusion coefficients in disordered systems such as spin glass systems and living cells are highly heterogeneous  and may change over time.  
Such a time-dependent and spatially heterogeneous environment results in irreproducibility 
of single-particle-tracking measurements. Irreproducibility of time-averaged observables has been theoretically studied 
in the context of weak ergodicity breaking in stochastic processes. 
Here, we provide rigorous descriptions of equilibrium and non-equilibrium diffusion processes for the 
annealed transit time model, which is a heterogeneous diffusion model in living cells. 
%These equations can be represented by recurrence time distributions in renewal processes. Solving the equations, 
We give analytical solutions for the mean square displacement (MSD) and the relative standard deviation of the time-averaged MSD 
for equilibrium and non-equilibrium situations. We find that
the time-averaged MSD grows linearly with time and that the diffusion coefficients are intrinsically random 
in non-equilibrium situations. Our findings pave the way for a theoretical understanding of distributional behavior of the
 diffusion coefficients in disordered systems. 
\end{abstract}

\noindent{\it Keywords\/}: anomalous diffusion, ergodicity, non-equilibrium processes

%\maketitle

\section{Introduction}
Transporting biological molecules 
%including  water molecules around cell membranes 
in living cells plays a key role in biochemical interactions, 
transmembrane signaling, and efficient reactions.  In single-particle tracking (SPT), the motion of proteins or lipids is tracked to  determine directly the 
diffusivity and to understand the biological role of diffusivity. Therefore, it is expected that 
SPT experiments will provide new insight into molecular transport in living cells.
In fact, many SPT experiments reveal anomalous dynamics such as subdiffusion, aging, fluctuating diffusivity, and heterogeneous environments in 
living cells \cite{Golding2006, Jeon2011, Weigel2011, Tabei2013, Hofling2013, Manzo2015}.  
%Molecular dynamics (MD) simulation is also a powerful tool for tracking individual molecules. In particular, the water retardation, i.e., slow diffusivity of 
%water molecules on the membrane surface, was clearly observed by MD simulations \cite{Yamamoto2013, Yamamoto2014}. 

Mean square displacement (MSD) is the most popular observable for characterizing the diffusivity of particles. 
There are two different averaging procedures for calculating the MSD. One is the ensemble average, and the other is the time 
average. The time-averaged MSD is defined as
\begin{equation}
 \label{tamsd_definition}
  \overline{\delta^{2}(\Delta;t)} \equiv
  \frac{1}{t - \Delta} \int_{0}^{t - \Delta} dt' \,
  [\bm{r}(t' + \Delta) - \bm{r}(t')]^{2},
\end{equation}
where ${\bm r}(t')$ is the position of a particle at time $t'$ tracked by the SPT experiments and $t$ is the total measurement time. 
  In stationary stochastic processes, these two averages are equivalent with the aid of the 
law of large numbers. This equivalence is one of the properties of ergodicity. While ergodicity is a concept 
in dynamical systems, an observable in a stochastic system is called ergodic if the time averages of the observable for different realizations 
converge uniquely to the ensemble average in equilibrium. 
This property ensures the reproducibility of measurements in experiments \cite{Barkai2012}:
 long SPT measurements give the same result under the same experimental setup. However, it was reported in SPT experiments in living cells
 that this reproducibility breaks down \cite{Jeon2011, Weigel2011, Tabei2013, Manzo2015, Graneli2006, Wang2006}, where the time-averaged 
MSD for a fixed $\Delta$ does not converge to a constant but fluctuates randomly across in realizations (random diffusion coefficient). 
Further, other experiments also reveal that time averages of observables such as occupation time and intensity of fluorescence fail to converge 
to a constant in some non-equilibrium systems \cite{Brokmann2003, Stefani2009, Takeuchi2015}. 
While there are several distributional limit theorems related to distributional behaviors of 
time averages in probability theory \cite{Darling1957, Lamperti1958, Dynkin1961}, little is known 
about the relationship between the stochastic models used in probability theory and the systems in experiments. 
Therefore, a theoretical foundation of irreproducibility is an important and challenging problem in statistical physics. 

Ergodicity gives a mathematical guarantee that time averages are equal to the ensemble average, i.e., it ensures reproducibility 
\cite{Birkhoff1931}. 
%Infinite ergodic theory has attract interest in  \cite{Lutz2013, Akimoto2010, Akimoto2012}.
Mathematically, infinite ergodic theory generalizes the concept of ergodicity, and 
states that time-averaged observables remain random even in the long-time limit \cite{Aaronson1997, TZ2006}.
%, and the distribution function depends on the system as well as a class of the observation function  
%\cite{Aaronson1981, Aaronson1997, Thaler1998, Akimoto2008, Akimoto2015}. 
Thus, it is expected that infinite ergodic theory will play a fundamental role in understanding 
random transport coefficients observed in SPT trajectories \cite{Akimoto2010, Akimoto2012, Lutz2013}. 
However, ergodicity in stochastic processes has been studied in a different way. If there is a 
highly stuck region in phase space, a particle cannot explore the whole phase space due to the trapping in the stuck region. Such a situation is called 
weak ergodicity breaking \cite{Bouchaud1992}. When a system shows weak ergodicity breaking, a time-averaged observable  does not 
converge to a constant even when the measurement time goes to infinity \cite{Lubelski2008, He2008, Metzler2014}. 
However, distributional behavior of time-averaged observables can be observed in homogeneous systems. In simple random walk, 
 the time-averaged occupation time that a random walker resides in positive region does not converge to a constant but converges in distribution, 
 known as the generalized arcsine law \cite{Feller1968}. 
Thus, some time-averaged observables in homogeneous environments do not converge to  constants but converge in distribution.
Such a time-averaged observable is not reproducible but has a distributional reproducibility because the distribution is universal 
in the sense that it does not depend on initial ensembles.
In stochastic models of anomalous diffusion, several distributional limit theorems for random diffusion coefficients 
have been studied to elucidate  irreproducibility \cite{Lubelski2008, He2008, Miyaguchi2011, Akimoto2013a, Metzler2014, Miyaguchi2015}. 
However, there are experimental results which cannot be explained by such stochastic models \cite{Manzo2015}. 
The goal of this paper is to fill the gap between experimentally observed irreproducibility and  
 distributional limit theorems in theoretical models. 

To consider the irreproducibility of the time-averaged MSD in diffusion in living cells, we investigate the annealed transit time model (ATTM) 
\cite{Massignan-Manzo-TorrenoPina-GarciaParajo-Lewenstein-Lapeyre-2014}, which has been shown to describe heterogeneous 
diffusion in living cells \cite{Manzo2015}. 
The authors of \cite{Massignan-Manzo-TorrenoPina-GarciaParajo-Lewenstein-Lapeyre-2014} 
show anomalous diffusion and aging of the time-averaged MSD. However, the distributional behavior of the time-averaged MSD remains an open problem. 
Moreover, the exact descriptions of the governing equations for the propagator have not been found so far. In this paper, we describe 
the equations rigorously and solve them analytically.  
Within this model, %similar as in \cite{Akimoto2013a, Akimoto2015}, 
we show that the time-averaged MSD remains 
random even in the long measurement times, i.e., the diffusion coefficients are irreproducible but have distributional reproducibility in the sense 
that the distribution of the time time-averaged MSD is universal.
%and the distribution function is different from that in CTRW. 

\section{Model}
In living cells, diffusivity strongly depends on space as well as time, that is, it is heterogeneous diffusion. One of the simplest models describing 
such a heterogeneous 
diffusion process is the Langevin equation with fluctuating diffusivity  \cite{Uneyama2015, Akimoto2016},
\begin{equation}
 \frac{d\bm{r}(t)}{dt} = \sqrt{2 D(t)} \bm{w}(t),
 \label{LEFD}
\end{equation}
where $\bm{r}(t)$ is the $n$-dimensional position of a particle at time $t$ and 
$D(t)$ is a stochastic process. Such a fluctuating diffusivity results from a fluctuating medium driven by fluctuations of friction or temperature
 \cite{Rozenfeld1998, Luczka2000, Beck2003},  diffusion in two-layer medium \cite{Luczka1995, Akimoto2015b}, or fluctuations of a diffusing particle's shape.
In \cite{Miyaguchi2016,Akimoto2016}, dichotomous processes are used for $D(t)$ to investigate effects of the underlying 
stochastic process $D(t)$ on physical features of the time-averaged MSD. 
To consider heterogeneous diffusion in living cells,  we have to model the stochastic process 
of $D(t)$. In a previous study, the ATTM was proposed for modeling  heterogeneous diffusion in living cells, 
where the diffusion coefficient is constant for a random sojourn time and the constant depends on the sojourn time 
\cite{Massignan-Manzo-TorrenoPina-GarciaParajo-Lewenstein-Lapeyre-2014}. When we consider quenched environment with 
heterogeneous local diffusivities, sojourn times in slow and high diffusive regions will imply long and short times, respectively. Thus, 
it is physically natural to assume that the sojourn time is inversely coupled to the diffusion coefficient. Moreover, it is important to consider 
the annealed model like Eq.~(\ref{LEFD}) because the annealed framework enables us to treat analytical calculations, heterogeneous 
environments in cells may not provide quenched environments, and the annealed model is considered to be a good approximation 
for the quenched model. 
In this paper, we assume that the diffusion coefficient is coupled to 
the sojourn time, i.e., $D_\tau =\tau^{\sigma -1}$ ($0<\sigma <1$) as in  \cite{Massignan-Manzo-TorrenoPina-GarciaParajo-Lewenstein-Lapeyre-2014}. 
In non-equilibrium situations, the probability density function (PDF) $\rho(\tau)$ of the sojourn time follows a power-law with no finite mean: 
\begin{equation}
\rho(\tau) \sim \frac{c}{|\Gamma(-\alpha)|}  \tau^{-1-\alpha}\quad (\tau \to \infty),
\end{equation}
where $c$ is a scale parameter.  A power-law sojourn-time distribution is observed in super-cooled liquids 
\cite{Helfferich2014} and can be derived with $\alpha=0.5$ in the first passage time. 
The mean sojourn time diverges for $\alpha \leq 1$, which means that there is no finite characteristic
time in the process. In other words, this process is an intrinsic non-equilibrium process.

 \begin{figure*}
\begin{center}
\includegraphics{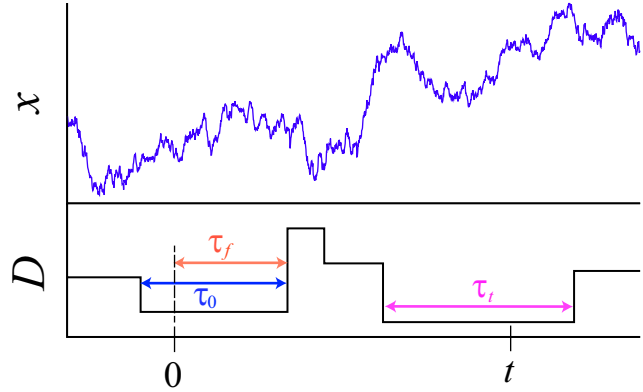}
\caption{Time series $x(t)$ and the underlying diffusion process $D(t)$ ($\sigma=0.1$). The first sojourn time ($\tau_0$), 
the forward recurrence time ($\tau_f$), and the sojourn time at time $t$ ($\tau_t$) are shown in the lower panel.
}
\end{center}
\label{traj_diff}
\end{figure*}
 
\section{Recurrence time distributions}

Here we provide several recurrence time distributions studied in renewal theory \cite{Cox, God2001}.
As shown in Fig.~\ref{traj_diff}, the underlying diffusion process at time $t=0$, $D(0)$, is determined by the first sojourn time 
$\tau_0$ and not by the forward recurrence time $\tau_f$.  Therefore, one must consider recurrence time distributions to describe the exact equations 
for the propagator in the equilibrium situation. By the same technique in \cite{God2001}, 
the Laplace transform of the joint PDF $f(\tau_t, \tau_f; t)$ of 
the sojourn time at time $t$, $\tau_t \equiv \tau_{N_t}$, and the forward recurrence time 
at time $t$, $\tau_f \equiv t_{N_t +1} - t$, is given by
\begin{eqnarray}
%\fl
\hat{f}(k, u; s) \equiv \int_0^\infty \int_0^\infty \int_0^\infty d\tau_t d\tau_f dt e^{-k \tau_t} e^{-u\tau_f} e^{-st} f(\tau_t,\tau_f;t)\nonumber\\
= \sum_{n=0}^\infty \left\langle e^{-k\tau_{n+1} - ut_{n+1}} \int_{t_n}^{t_{n+1}} dt e^{-(s-u)t} \right\rangle 
%\nonumber\\
= \frac{\hat{\rho}(k+s) - \hat{\rho}(k+u)}{u-s} \frac{1}{1-\hat{\rho}(s)},
\end{eqnarray}
where $N_t$ is the number of changes of states until time $t$, $t_k$ is the time when the $k$th change of states occurs, 
 and $\hat{\rho}(s)$ is the Laplace transform of $\rho(\tau)$.
In equilibrium process, the system started at $t=-\infty$, and we start to observe from $t=0$ \cite{Cox}. Thus, we have 
the double Laplace transform of the joint PDF $f_{\rm eq}(\tau_t,\tau_f)$ of $\tau_t$ and $\tau_f$ in equilibrium process:
\begin{equation}
\hat{f}_{\rm eq}(k, u) = \lim_{s\to 0} s\hat{f}(k, u; s) =\frac{\hat{\rho}(k) - \hat{\rho}(k+u)}{\langle \tau\rangle u},
\end{equation}
where $\langle \tau \rangle \equiv \int_0^\infty \rho(\tau)d\tau$ is the mean sojourn time. 
The inverse Laplace transform with respect to $k$ and $u$ yields
\begin{equation}
f_{\rm eq}(\tau_t, \tau_f) =  \frac{\rho(\tau_t) \theta(\tau_t-\tau_f)}{\langle \tau \rangle},
\label{joint_PDF_taut_tauf}
\end{equation}
where $\theta(t)=1$ if $t>0$ and $\theta(t)=0$ otherwise. 
Integrating Eq.~(\ref{joint_PDF_taut_tauf}) with respect to $\tau_f$ yields the PDF $\rho_{\rm eq}(\tau)$ of the sojourn times 
at $t=0$ in equilibrium ($\tau_t$ for $t\to \infty$):
\begin{equation}
\rho_{\rm eq}(\tau) = \int_0^\infty f_{\rm eq} (\tau,\tau_f) d\tau_f= \frac{\tau \rho(\tau)}{\langle \tau\rangle}.
\end{equation}
%which has the mean $2\langle \tau \rangle$, and the PDF of the forward recurrence time is given by 
%{\color{red}$\rho_0(\tau)=\int_\tau^\infty \rho(\tau)d\tau/\langle \tau \rangle$. }
The mean and the second moments of the initial diffusion coefficient $D(0)$ in equilibrium can be calculated as  
$
\langle D(0) \rangle_{\rm eq} \equiv \int_0^\infty d\tau D_\tau \rho_{\rm eq}(\tau) %=\frac{\Gamma(1+\sigma)}{\langle \tau \rangle^{1-\sigma}}
$
and
$
\langle D(0)^2 \rangle_{\rm eq} \equiv \int_0^\infty d\tau D_\tau^2 \rho_{\rm eq}(\tau)%=  \frac{\Gamma(2\sigma)}{\langle \tau \rangle^{2(1-\sigma)}}
$, and are assumed to be finite in equilibrium processes.

\section{General framework}

Let $P({\bm r},t)$ be the PDF of the position ${\bm r} = (r_1, \cdots, r_n)$ at time $t$ 
and $Q({\bm r},t)$ be the PDF of the position ${\bm r}$ conditioned that the state of $D(t)$ changes at exactly 
 time $t$. We assume that $P({\bm r},0)=\delta({\bm r} )$. Hence, the PDFs satisfy the following generalized renewal equations:
%\begin{widetext}
\begin{eqnarray}
\fl
Q({\bm r},t) = \int_{-\infty}^\infty dr'_1\cdots \int_{-\infty}^\infty dr'_n \int_0^t dt'\psi({\bm r}',t') Q({\bm r}-{\bm r}',t-t')
+\psi_0 ({\bm r},t), \label{GRW_Q}\\
\fl
P({\bm r},t;\tau) = \int_0^t dt' \Psi({\bm r}',t';\tau) Q({\bm r}-{\bm r}',t-t') + \Psi_0 ({\bm r},t;\tau),
\label{GRW_P}
\end{eqnarray}
%\end{widetext}
where $P({\bm r},t;\tau)$ is the PDF $P({\bm r},t)$ conditioned that the sojourn at $t$, $\tau_t$, is given by $\tau$,
 $\psi({\bm r},t)$ is the joint PDF of the displacement ${\bm r}$ and the sojourn time $t$,  
$\psi_0 ({\bm r},t)$ is the joint PDF of the displacement ${\bm r}$ and the first sojourn time $t$, 
 $\Psi({\bm r},t;\tau)$ is the joint PDF of the displacement ${\bm r}$, the time elapsed $t$ from $t_{N_t}$, and 
 the last sojourn time (the sojourn time at $t$)
 $\tau$. Finally, the PDF $P({\bm r},t)$ is 
\begin{equation}
P({\bm r},t) = \int_0^\infty d\tau P({\bm r},t;\tau),
\end{equation}
where $\Psi_0 ({\bm r},t;\tau)$ is the joint PDF of the displacement ${\bm r}$ and the time elapsed $t$, and the sojourn time $\tau$, and
 there is no renewal during $t$. 

In ATTM, the joint PDF $\psi({\bm r},t)$ is given by $\psi({\bm r},t)=\phi({\bm r},t)\rho(t)$, where $\phi({\bm r},t)$ is a Gaussian propagator with diffusion coefficient 
$D_t=t^{\sigma-1}$:
\begin{equation}
\phi({\bm r},t)=\frac{1}{2\sqrt{n\pi D_t t}} \exp \left( \frac{-{\bm r}^2}{4n D_t t}\right).
\end{equation}
Moreover, the joint PDF $\Psi({\bm r},t;\tau)$ is given by $\Psi({\bm r},t; \tau)= \rho(\tau)\phi({\bm r},t;\tau) \theta (\tau-t)$, 
where $\phi({\bm r},t;\tau)$ is a Gaussian propagator with diffusion coefficient 
$D_\tau=\tau^{\sigma-1}$:
\begin{equation}
\phi({\bm r},t;\tau)=\frac{1}{2\sqrt{n\pi D_\tau t}} \exp \left( \frac{-{\bm r}^2}{4 nD_\tau t}\right),
\end{equation}
 and the joint PDF $\Psi_0 ({\bm r},t;\tau)$ is given by 
$\Psi_0 ({\bm r},t;\tau) = \phi({\bm r},t;\tau) \int_t^\infty f_{\rm eq} (\tau,\tau_f) d\tau_f$. 
By the Fourier-Laplace transform, we have from Eqs.~(\ref{GRW_Q}) and (\ref{GRW_P})
\begin{eqnarray}
\hat{P}({\bm k},s) =  \frac{ 1 + \hat{\psi}_0({\bm k},s)}{1-\hat{\psi}({\bm k},s)} \int_0^\infty d\tau \hat{\Psi}({\bm k},s;\tau)
 + \int_0^\infty d\tau \hat{\Psi}_0 ({\bm k},s;\tau). 
\label{FLT_propagator}
\end{eqnarray} 
This is the exact representation of the Fourier-Laplace transform of the propagator, which is a generalization of the random walk 
framework \cite{Montroll1965, Scher1973, Shlesinger1982, Shlesinger1987, Akimoto2014}.
%where
%$\hat{P}({\bm k},s) \equiv \int_{-\infty}^\infty d{\bm r} \int_0^\infty dt  e^{-st-{\bm k}\cdot{\bm r}}P({\bm r},t)$,
%$\hat{\psi}({\bm k},s) \equiv \int_{-\infty}^\infty d{\bm r} \int_0^\infty dt  e^{-st-{\bm k}\cdot{\bm r}}\psi({\bm r},t)$,
%and
%$\hat{\Psi}({\bm k},s;\tau) \equiv \int_{-\infty}^\infty d{\bm r} \int_0^\infty dt  e^{-st-{\bm k}\cdot{\bm r}}\Psi({\bm r},t;\tau)$.

%\subsection{Moments of TAMSD}
Next, we derive moments of the time-averaged MSD. 
For $\Delta \ll t$, we approximate the time-averaged MSD as
\begin{equation}
\overline{\delta^{2}(\Delta;t)} \sim  \frac{2n}{t}  \left(\sum_{i=0}^{N_t} D_{\tau_i} (\tau_i)\tau_i  
+  (t-t_{N_t})  D_{\tau_{N_t+1}}(\tau_{N_t +1})\right) \Delta,
\label{tmsd}
\end{equation}
where $N_t$ is the number of changes of states until time $t$, $\tau_i$ is the $i$th sojourn time, $t_i\equiv \tau_1 + \cdots +\tau_i$, and 
$D_{\tau} (\tau)$ is the time-averaged diffusion coefficient under the diffusion coefficient $D_\tau$:
$
D_{\tau} (\tau) \equiv \int_0^\tau \{{\bm r}(t'+\Delta) - {\bm r}(t')\}^2dt'/ (\tau \Delta).
$
We further assume that $D_\tau(\tau)=D_\tau$: 
\begin{equation}
\frac{t\overline{\delta^{2}(\Delta;t)}}{2n\Delta}\sim 
Z(t)\equiv  \sum_{i=0}^{N_t} D_{\tau_i}\tau_i + D_{\tau_{N_t+1}} (t-t_{N_t}).
\end{equation}
This assumption is considered to be valid in the asymptotic limit for $t$ \cite{Akimoto2016}. 
Let $P_D(z,t)$ be the PDF of  $Z(t)$ 
and $Q_D(x,t)$ be the PDF of $Z(t)$ where a renewal occurs at exactly  time $t$. We can write 
the generalized renewal equation for $Z(t)$:  
%\begin{widetext}
\begin{eqnarray}
Q_D(z,t) &= &\int_{0}^z dz' \int_0^t dt'\psi_D(z',t') Q_D(z-z',t-t')  + \psi_D^0 (z,t), \label{GRE_QD}\\
P_D(z,t;\tau) &=& \int_0^t dt' \Psi_D(z',t';\tau) Q_D(z-z',t-t') +\Psi_D^0 (z,t;\tau), \label{GRE_PD}
\end{eqnarray}
%\end{widetext}
where $\psi_D(z,t)$ is the joint PDF of $Z(t)$ and the time elapsed $t$,  i.e., $\psi_D(z,t)=\rho(t) \delta(z-t^{\sigma})$, 
$\psi_D^0(z,t)$ is the joint PDF of $Z(t)$ and the first renewal at $t$, 
% i.e., $\psi_D^0(z,t)= \int_0^\infty d\tau f_{\rm eq}(\tau, t) \delta(z-\tau^{\sigma-1}t)$,
$\Psi_D(z,t;\tau)$ is the joint PDF of the displacement $Z$, the time elapsed $t$, and the last sojourn time 
given by $\tau$, i.e., $\Psi_D(z,t;\tau)= \rho(\tau) \delta(z-\tau^{\sigma-1} t) \theta(\tau-t)$, and 
$\Psi_D^0(z,t;\tau)$ is the joint PDF of the displacement $Z$, the time elapsed $t$, and the last sojourn time $\tau$.
%, i.e., $\Psi_D^0 (z,t;\tau)= \int_t^\infty dt' f_{\rm eq}(\tau, t') \delta(z-\tau^{\sigma-1} t) \theta(\tau-t)$. 
Finally, $P_D(z,t)$ can be obtained: 
\begin{equation}
P_D(z,t) = \int_0^\infty d \tau P_D(z,t;\tau).
\end{equation}
By the double Laplace transform, we have 
\begin{equation}
\hat{P}_D(k,s)=  \frac{\hat{\psi}_D^0 (k,s) \int_0^\infty \hat{\Psi}_D(k,s;\tau) d\tau}{1-\hat{\psi}_D(k,s)}
+ \int_0^\infty \hat{\Psi}^0_D(k,s;\tau) d\tau.
\end{equation} 

\section{Equilibrium and Non-equilibrium Processes}

\subsection{Normal diffusion and fluctuation of the time-averaged MSD in equilibrium processes}
In equilibrium processes, the PDFs related to the first recurrence times are given by
$\psi_0 ({\bm r},t) = \int_0^\infty d\tau f_{\rm eq}(\tau,t) \phi ({\bm r},t;\tau) $ and
$\Psi_0 ({\bm r},t;\tau) = \phi ({\bm r},t;\tau) \int_t^\infty dt' f_{\rm eq}(\tau,t')$. 
Substituting these into Eq.~(\ref{FLT_propagator}), we obtain  the Laplace transform of the MSD: 
$\langle {\bm r}(s)^2\rangle_{\rm eq} = \left. \sum_{i=1}^n \frac{\partial^2 \hat{P}}{\partial {\bm k}_i^2} \right|_{{\bm k}={\bm 0}}
= \frac{2n}{s^2} \int_0^\infty d\tau \frac{\rho(\tau)}{\langle \tau\rangle}D_\tau \tau$.
It follows that the MSD grows linearly with time in equilibrium processes:
\begin{equation}
\langle {\bm r}(t)^2 \rangle_{\rm eq} = 2n \langle D(0) \rangle_{\rm eq}t. 
\end{equation}

\begin{figure}
\begin{center}
\includegraphics[width=.6\linewidth, angle=0]{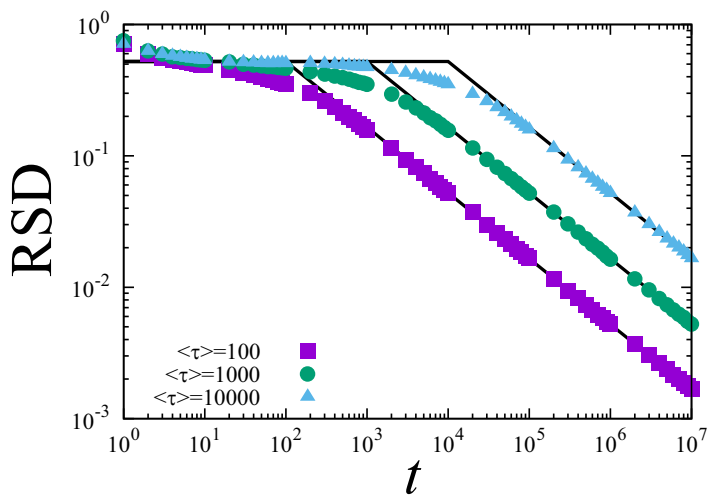}
\caption{Relative standard deviation of the time-averaged MSD as a function of the measurement time ($\sigma=0.8$). 
We used exponential distributions with different relaxation times for the sojourn time distribution. 
The lines represent the theory, i.e., Eqs.~(\ref{theory_rsd_asym_exp}) and (\ref{theory_rsd_flat_exp}), while symbols show the results of numerical simulations.}
\label{rsd_exp}
\end{center}
\end{figure}

Moreover, the PDFs related to the first recurrence times in $Z(t)$ are given by 
$\psi_D^0(z,t)= \int_0^\infty d\tau f_{\rm eq}(\tau, t) \delta(z-\tau^{\sigma-1}t)$ and 
$\Psi_D^0 (z,t;\tau)= \int_t^\infty dt' f_{\rm eq}(\tau, t') \delta(z-\tau^{\sigma-1} t) \theta(\tau-t)$.
The Laplace transform of $\langle Z(t)\rangle$, denoted by $\langle \hat{Z}(s)\rangle$, is given by
$\langle \hat{Z}(s)\rangle = - \left. \frac{\partial \hat{P}_D(k,s)}{\partial k}\right|_{k=0} 
= \int_0^\infty d\tau \frac{\rho(\tau)\tau^\sigma}{\langle \tau \rangle s^2}$.
The inverse Laplace transform reads $Z(t) = \langle D(0) \rangle_{\rm eq}t$. Hence, 
$\langle \overline{\delta^2 (\Delta;t)}\rangle_{\rm eq} = 2n \langle D(0) \rangle_{\rm eq} \Delta$. 
To characterize the irreproducibility of the time-averaged MSD, we calculate the relative standard deviation (RSD) studied in 
several diffusion processes
\cite{He2008, Deng2009, Akimoto2011}:
\begin{equation}
\Sigma(t;\Delta) \equiv \frac{\sqrt{\langle [\overline{\delta^{2}(\Delta;t)} -
 \langle \overline{\delta^{2}(\Delta;t)} \rangle]^{2} \rangle}}
 {\langle \overline{\delta^{2}(\Delta;t)} \rangle} .
\end{equation}
We note that the RSD is independent of $\Delta$ because the time-averaged MSD depends linearly on
$\Delta$ in the ATTM [see Eq.~(\ref{tmsd})].
If the time-averaged MSD is reproducible, then the RSD approaches zero as the measurement time $t$ goes to infinity. 
It is important to note the RSD extracts a characteristic time from the system even when the time-averaged MSD is reproducible 
\cite{Miyaguchi2011a, Uneyama2012, Akimoto2015b}. In particular, as will be shown below, a crossover time in the RSD 
is related to a characteristic time of fluctuating diffusivity if the instantaneous diffusivity changes over time.
Obtaining the Laplace transform of $\langle Z^2(t)\rangle$ and inverting it, we have the asymptotic behavior of the squared RSD:
\begin{equation}
\fl
\Sigma^2(t;\Delta) %\sim \frac{\langle Z(t)^2 \rangle_{\rm eq} - \langle Z(t) \rangle_{\rm eq}^2}{\langle Z(t) \rangle_{\rm eq}^2}
\sim \frac{1}{t} \left( \frac{\langle \tau^2 \rangle}{\langle \tau \rangle } - \frac{2\int_0^\infty d\tau \rho(\tau)\tau^{\sigma+1}}
 {\int_0^\infty d\tau \rho(\tau)\tau^\sigma } +  \frac{\langle \tau \rangle\int_0^\infty d\tau \rho(\tau)\tau^{2\sigma}}
 {\left(\int_0^\infty d\tau \rho(\tau)\tau^\sigma \right)^2}\right)
 \quad (t\to \infty).
\end{equation}
When $\rho(\tau)$ is the exponential distribution, the asymptotic behavior of the squared RSD decays as
\begin{equation}
\Sigma^2(t;\Delta) 
\sim \frac{\langle \tau \rangle}{t} \left(\frac{\Gamma(2\sigma + 1)}{\Gamma(\sigma + 1)^2} - 2\sigma \right)\quad (t\to \infty).
\label{theory_rsd_asym_exp}
\end{equation}
Thus, the RSD becomes zero when the measurement time goes to infinity. In other words, the time-averaged MSD 
is reproducible in the long-time measurements.
On the other hand, for measurement times that are small compared with the characteristic time $\tau_c$, the RSD does not decay:
\begin{equation}
\Sigma(t;\Delta) \cong \frac{\sqrt{\langle D(0)^2 \rangle_{\rm eq} - \langle D(0) \rangle_{\rm eq}^2}}{\sqrt{\langle D(0) \rangle}_{\rm eq}}
=\sqrt{\frac{\Gamma(2\sigma) }{\Gamma(1+\sigma)^2} - 1}  \quad (t \ll \tau_c).
\label{theory_rsd_flat_exp}
\end{equation}
Thus, there is a transition  from constant RSD (irreproducible) to reproducible behavior, and the crossover time is related to 
a characteristic time like the mean sojourn time (see Fig.~\ref{rsd_exp}). The crossover time provides useful information 
on a characteristic time of fluctuating diffusivity, which has not been known so far.

%In the case of a power-law PDF, $\rho(\tau) =\alpha \tau^{-1-\alpha}$ ($\tau \geq 1$),
%\begin{equation}
%\Sigma^2(t;\Delta) 
%\sim \frac{1}{t} \left( \frac{\alpha -1}{\alpha -2} + \frac{(\alpha - \sigma)^2}{(\alpha -1)(\alpha -2 \sigma)} 
%- \frac{2(\alpha -\sigma)}{\alpha -\sigma -1} \right) \quad (t\to \infty),
%\end{equation}
%and
%\begin{equation}
%\Sigma(t;\Delta) \cong \frac{\sqrt{\langle D(0)^2 \rangle_{\rm eq} - \langle D(0) \rangle_{\rm eq}^2}}{\sqrt{\langle D(0) \rangle}_{\rm eq}}
%=\sqrt{\frac{(\alpha -\sigma)^2 }{(\alpha-1)(\alpha -2\sigma +1)} - 1}  \quad (t \ll \tau_c).
%\label{theory_rsd_flat_pow}
%\end{equation}

\subsection{Anomalous diffusion, aging, and distributional reproducibility in non-equilibrium processes}

\begin{figure}
\begin{center}
\includegraphics[width=.6\linewidth, angle=0]{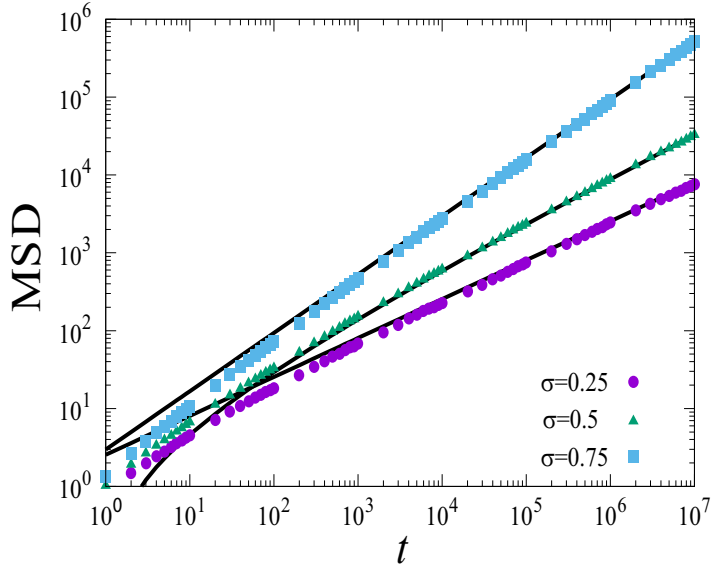}
\caption{Mean square displacements for different $\sigma$ ($\alpha=0.5$ and $n=1$). 
The lines represent the theory~(\ref{theory_msd_noneq}), while the symbols show the results of numerical simulations. 
Note that there are no fitting parameters.}
\label{msd_noneq}
\end{center}
\end{figure}

Here, we assume that the PDF of the sojourn time follows a power-law distribution $\rho(\tau)$ with exponent $\alpha<1$. 
%and the diffusion coefficient in a state is determined by $\tau^{\sigma-1}$ ($0<\sigma <{\color{red}1}$), where $\tau$ is a sojourn time in this state.
Because there is no equilibrium distribution for the forward recurrence time, this stochastic process is an intrinsic non-equilibrium process. 
If $\sigma > \alpha$, we note that the average of $2nD_t t$ with respect to the sojourn time $t$, i.e., the MSD during times when the state 
does not change, diverges because
$\langle D_t t \rangle\equiv  \int_0^\infty dt \rho(t) D_t t= \infty$. %\subsection{Ensemble-averaged MSD}
We consider a non-equilibrium situation in which the first renewal occurs at time $t=0$. In this case, the generalized renewal equation 
is given by setting $\psi_0 ({\bm r},t)=0$ and $\Psi_0 ({\bm r},t;\tau)=0$ in Eqs. (\ref{GRW_Q}) and (\ref{GRW_P}).
%\begin{eqnarray}
%Q({\bm r},t) &= \int_{-\infty}^\infty d{\bm r}_1'\cdots \int_{-\infty}^\infty d{\bm r}_n' \int_0^t dt'\psi({\bm r}',t') Q({\bm r}-{\bm r}',t-t'), \label{Qrt_noneq} \\
%P({\bm r},t;\tau) &= \int_0^t dt' \Psi({\bm r}',t';\tau) Q({\bm r}-{\bm r}',t-t'), \label{Prt_noneq}
%\end{eqnarray}
Using the Laplace analysis as in the equilibrium case, we have
the asymptotic behavior of $\langle {\bm r}(t)^2 \rangle$ for $t\to \infty$:
\if0
\begin{equation}
\langle {\bm r}(s)^2\rangle \sim \left\{
\begin{array}{ll}
\frac{2n\Gamma(\sigma - \alpha)}{|\Gamma(-\alpha)|(1+\alpha-\sigma)} \frac{1}{s^{1+\sigma}} \quad &(\sigma>\alpha),\\
\\
\frac{2n}{|\Gamma(-\alpha)|} \frac{\ln(1/s)}{s^{1+\alpha}} + \frac{2n}{|\Gamma(-\alpha)| s^{1+\alpha}}\quad &(\sigma=\alpha),\\
\\
\frac{\langle 2nD_t t\rangle}{c} \frac{1}{s^{1+\alpha}} + \frac{2n\Gamma(1 - \alpha +\sigma)}{|\Gamma(-\alpha)|(1+\alpha-\sigma)} \frac{1}{s^{1+\sigma}}
\quad &(\sigma<\alpha).
\end{array}
\right.
\end{equation}
The inverse Laplace transform reads
\fi
\begin{equation}
\langle {\bm r}(t)^2\rangle \sim \left\{
\begin{array}{ll}
\frac{2n\Gamma(\sigma -\alpha)}{|\Gamma(-\alpha)|(1+\alpha-\sigma) \Gamma(1+\sigma)} t^{\sigma} \quad &(\sigma>\alpha),\\
\\
\frac{2n}{|\Gamma(-\alpha)|\Gamma(1+\alpha)} t^\alpha \ln t %\left(1 + \frac{1}{\ln t}\right)\quad 
&(\sigma=\alpha),\\
\\
\frac{2n\langle D_t t\rangle}{c \Gamma(1+\alpha)}t^\alpha 
%\left(1+ \frac{c\Gamma(1+\alpha)\Gamma(1 -\alpha+\sigma)}{\langle D_t t\rangle |\Gamma(-\alpha)|(1+\alpha-\sigma) \Gamma(1+\sigma)} \frac{1}{t^{\alpha-\sigma}}\right)
\quad &(\sigma<\alpha).
\end{array}
\right.
\label{theory_msd_noneq}
\end{equation}
Our theory provides the exact form of the MSD in the asymptotic limit, which 
matches perfectly  with the results of numerical simulations without fitting the parameters (see Fig.~\ref{msd_noneq}).
The exponent of subdiffusion is the same as that previously obtained (note that our notations are described by 
$\alpha=\sigma/\gamma$ and $\sigma=1-1/\gamma$ in their notations) \cite{Massignan-Manzo-TorrenoPina-GarciaParajo-Lewenstein-Lapeyre-2014}.

\begin{figure}
\begin{center}
\includegraphics[width=.6\linewidth, angle=0]{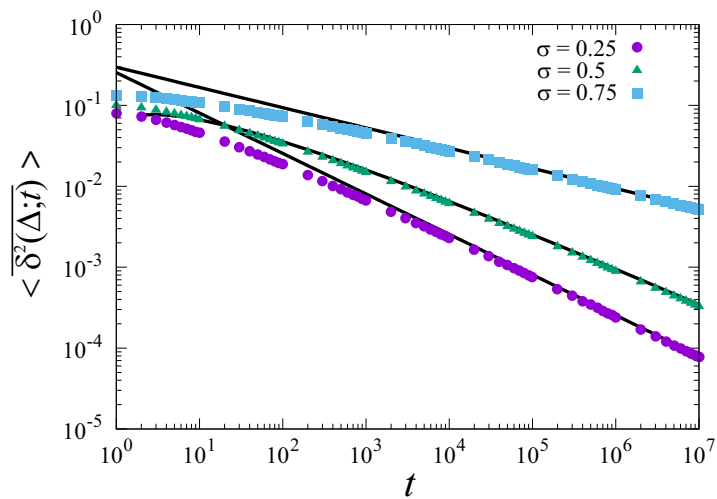}
\caption{Ensemble average of the time-averaged MSD as a function of the measurement time $t$ for different $\sigma$ ($\alpha=0.5$ and $n=1$). 
The lines represent the theory~(\ref{theory_tmsd_aging}), while the symbols show the results of numerical simulations. 
Note that there are no fitting parameters.}
\label{tmsd_aging}
\end{center}
\end{figure}

%\subsection{Moments of TAMSD}
Next, we derive the ensemble average of the time-averaged MSD. The generalized renewal equation for $Z(t)$ is given by setting 
 $\psi_D^0 (z,t)=\delta (z) \delta(t)$ and $\Psi_D^0 (z,t;\tau)=0$ in Eqs.~(\ref{GRE_QD}) and (\ref{GRE_PD}). 
%Because the mean sojourn time of state $D(t)$ is infinite in this case, $\Delta (\ll t)$ is much smaller than a characteristic time of $D(t)$. 
%For $\Delta \ll t$,  we use the same approximation as in equilibrium.
%Let $P_D(z,t)$ be the PDF of  $Z(t)$ 
%and $Q(x,t)$ be the PDF of $Z(t)$ where a renewal occurs exactly at time $t$, 
%\begin{widetext}
%\begin{eqnarray}
%Q_D(z,t) &=\delta(z)\delta(t) + \int_{0}^z dz' \int_0^t dt'\psi_D(z',t') Q_D(z-z',t-t'),\\
%P_D(z,t;\tau) &= \int_0^t dt' \Psi_D(z',t';\tau) Q_D(z-z',t-t'),
%\end{eqnarray}
%\end{widetext}
%where $\psi_D(z,t)$ is the joint PDF of $Z(t)$ and time elapse $t$,  i.e., $\psi_D(z,t)=\rho(t) \delta(z-t^{\sigma})$, 
% $\Psi_D(z,t;\tau)$ is the joint PDF of displacement $Z$ and time elapse $t$ under the condition that the last sojourn time 
%is given by $\tau$, i.e., $\Psi_D(z,t;\tau)=\delta(z-\tau^{\sigma-1} t) \theta(\tau-t)$. Finally, we have
%\begin{equation}
%P_D(z,t) = \int_0^\infty \rho(\tau) P_D(z,t;\tau).
%\end{equation}
%By the double Laplace transform, we have 
%\begin{equation}
%\hat{P}_D(k,s)=  \frac{\int_0^\infty \rho(\tau) \hat{\Psi}_D(k,s;\tau) d\tau}{1-\hat{\psi}_D(k,s)}.
%\end{equation} 
Using the Laplace analysis on $P_D(z,t)$ as in the equilibrium case and using the relation $\langle \overline{\delta^{2}(\Delta;t)} \rangle \sim 
2n\Delta \langle Z(t) \rangle/t$, we have 
%\begin{widetext}
\begin{equation}
\langle \overline{\delta^{2}(\Delta;t)} \rangle \sim \left\{
\begin{array}{ll}
 \frac{2n\Delta \Gamma(\sigma -\alpha)}{|\Gamma(-\alpha)|(1+\alpha-\sigma) \Gamma(1+\sigma)} t^{\sigma-1} \quad &(\sigma>\alpha),\\
 \\
 \frac{2n\Delta}{|\Gamma(-\alpha)|\Gamma(1+\alpha)} t^{\alpha-1} \ln t \left( 1+ \frac{1}{\ln t}\right)&(\sigma=\alpha),\\
 \\
 \frac{2n\Delta \langle D_t t\rangle}{c \Gamma(1+\alpha)}t^{\alpha-1} %+ \frac{2n\Delta\Gamma(1-\alpha+\sigma )}{|\Gamma(-\alpha)|(1+\alpha-\sigma)} 
 %\frac{t^\sigma}{\Gamma(1+\sigma)}
 &(\sigma<\alpha).
\end{array}
\right.
\label{theory_tmsd_aging}
\end{equation}
Therefore, the ensemble average of the time-averaged MSD shows aging: $\langle \overline{\delta^{2}(\Delta;t)} \rangle \to 0$ ($t\to \infty$). 
Figure~\ref{tmsd_aging} shows that this aging behavior is clearly described by Eq.~(\ref{theory_tmsd_aging}). 
This exact form in the asymptotic limit has also been obtained for the first time.

\begin{figure}
\begin{center}
\includegraphics[width=.6\linewidth, angle=0]{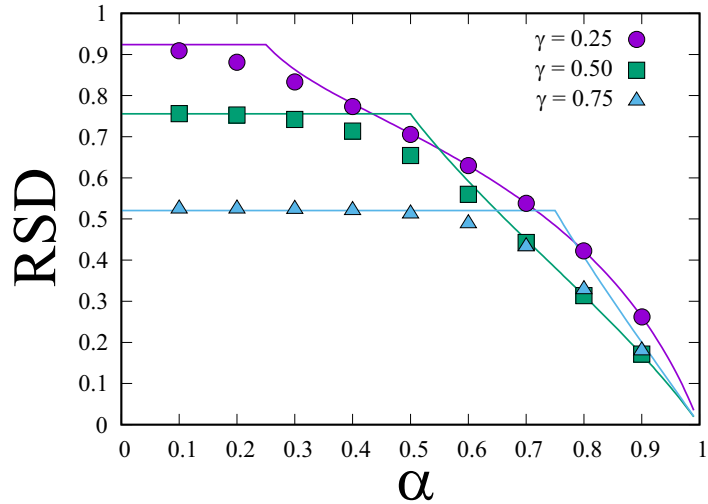}
\caption{Relative standard deviation of the time-averaged MSD as a function of $\alpha$ ($\sigma=0.25, 0.5$, and 0.75). 
In numerical simulations, we use $\Delta=0.1$ for calculating the RSD. 
The dashed lines represent the theory described by Eq.~(\ref{rsd_attm}), while the symbols show the results of 
numerical simulations.}
\label{rsd_crossover}
\end{center}
\end{figure}

Moreover, we obtain the second moment of $Z(t)$ (see Appendix.~A):
\begin{equation}
\langle \hat{Z}(t)^2\rangle \sim \left\{
\begin{array}{ll}
\left( \frac{2\Gamma(2\sigma-\alpha) }{|\Gamma(-\alpha)| (2+\alpha-2\sigma) }
+ \frac{2\Gamma(\sigma -\alpha)^2}{|\Gamma(-\alpha)|^2 (1+\alpha-\sigma)}\right)\frac{t^{2\sigma}}{\Gamma(1+2\sigma)} \quad 
&(\sigma > \alpha),\\
\\
\frac{ 2}{|\Gamma(-\alpha)|^2} \frac{(t^\alpha \ln t)^2}{\Gamma(2\alpha +1)}&(\sigma = \alpha),\\
\\
\frac{ 2\langle D_t t\rangle^2}{c^2} \frac{t^{2\alpha}}{\Gamma(2\alpha +1)}&(\sigma \leq \alpha).
\end{array}
\right.
\end{equation} 
It follows that the RSD is given by
\begin{equation}
\Sigma^2(t; \Delta) \sim \left\{
\begin{array}{ll}
\frac{2(1+\alpha-\sigma) \Gamma(1+\sigma)^2}{ \Gamma(1+2\sigma)} \left(
\frac{{(1+\alpha -\sigma)} \Gamma(2\sigma-\alpha) |\Gamma(-\alpha)|}{(2+\alpha-2\sigma)\Gamma(\sigma-\alpha)^2} +1 \right) -1\quad &(\sigma > \alpha),\\
\\
\frac{2\Gamma(1+\alpha)^2}{\Gamma(2\alpha +1)}-1&(\sigma \leq \alpha),
\end{array}
\right.
\label{rsd_attm}
\end{equation}
in the asymptotic limit of $t\to\infty$.
%\end{widetext}
Therefore, the RSD does not decay even in the long-time limit for the measurement time. The theory of the RSD has been 
confirmed in numerical simulations (see Fig.~\ref{rsd_crossover}).
This is direct evidence of  
irreproducibility. We note that the value of the RSD for $\sigma \leq \alpha$ is the same as that in a continuos-time 
random walk \cite{He2008}.  

Furthermore, we can show that all of the higher moments are given by
\begin{equation}
\langle \hat{Z}(t)^k\rangle \sim \left\{
\begin{array}{ll}
\frac{M_k(\alpha,\sigma)}{|\Gamma(-\alpha)|} \frac{t^{k(1-\sigma)}}{\Gamma(k+1-k\sigma)} \quad 
&(\sigma > \alpha)\\
\\
\frac{ k! \langle D_t t\rangle^k}{c^k} \frac{t^{k\alpha}}{\Gamma(k\alpha +1)}&(\sigma \leq \alpha),
\end{array}
\right.
\end{equation}
where the coefficient $M_k(\alpha,\sigma)$ is given in Appendix~A.
Because $Z(t)/t$ is the time-averaged diffusion coefficient, 
the distribution of the normalized time-averaged diffusion coefficient, 
$\overline{D_t}\equiv \overline{\delta^{2}(\Delta;t)}/\langle \overline{\delta^{2}(\Delta;t)}\rangle\sim Z(t)/\langle Z(t)\rangle$, 
does not converge to a delta function like ergodic observables but converge to a broad distribution. 
Moreover, the distribution of time-averaged diffusion coefficients 
obtained from single trajectories is universal in the sense that it does not depend on the initial conditions nor noise histories.
Therefore, time-averaged diffusion coefficient has a distributional reproducibility in the ATTM when the exponent is less than
one.

%%%%%%%%%%%%%%%%%%%%
%discussion
%%%%%%%%%%%%%%%%%%%%
\section{Discussion}
We have described rigorous equations for the propagator, Eqs.~(\ref{GRW_Q}) and (\ref{GRW_P}), and the time-averaged MSD, 
Eqs.~(\ref{GRE_QD}) and (\ref{GRE_PD}), in ATTM. By solving these equations, we have obtained exact solutions 
for the MSD and the moments of the time-averaged MSD. In equilibrium processes, we found a transition from irreproducible to reproducible 
behavior in the time-averaged MSD and extracted the characteristic time using the crossover time.  
However, the RSD does not decay at all in non-equilibrium processes. 
We have provided theoretical evidence for distributional reproducibility of the time-averaged MSD in heterogeneous environments. 
Distributional behaviors for the time-averaged MSD obtained here are closely related to the distributional limit theorem of a non-integrable 
observation function in infinite ergodic theory \cite{Akimoto2015}. This is because the moments obtained here are similar to those 
in \cite{Akimoto2015}. In other words, the distribution looks the same in shape. 

A quenched model, called the quenched radius model (QRM), was also considered in \cite{Massignan-Manzo-TorrenoPina-GarciaParajo-Lewenstein-Lapeyre-2014}.
By analogy to the relationship between the quenched trap model and the annealed model (continuous-time random walk) 
\cite{Machta1985, bouchaud90, Miyaguchi2011, Miyaguchi2015}, we conjecture that   
the exponent of the MSD as well as the moments of the time-averaged MSD in QRM will be the same as those in ATTM 
when the dimension is greater than two. 
% although the boundaries of patches in QRM may affect the results. 
However, it should be noted that the MSD and the moments of the time-averaged MSD differ when the 
dimension is less than two.  In fact, the subdiffusive exponent 
in the QRM is given by $2\alpha/(1+\alpha)$ in the one-dimensional case when the second moment of patch size does not diverge 
\cite{Massignan-Manzo-TorrenoPina-GarciaParajo-Lewenstein-Lapeyre-2014}. 
%Intuitively speaking, the MSD is given 
%by $\langle x(t)^2 \rangle = \langle r^2 \rangle \langle N_t\rangle$, where $r$ is a patch size and 
%$N_t$ is the number of different patches visited by a random walker until time $t$. 
%We conjecture that $\langle x(t)^2 \rangle \propto \langle N_t \rangle \propto t^{2\alpha/(1+\alpha)}$ if $ \langle r^2 \rangle<\infty$,
%$\langle x(t)^2 \rangle \propto t^\sigma$ otherwise. Moreover, we think that $\langle 2D_t t\rangle$ diverges when $2\alpha/(1+\alpha)\leq \sigma$. 
Thus, the ergodic properties for one-dimensional QRM will be different from those for ATTM, which is still an interesting open problem. 

\section*{ Acknowledgments}
%We thank T. Miyaguchi and T. Uneyama for discussions on these issues. 
T. A. was partially supported by a Grant-in-Aid for Young Scientists (B) (26800204). 
%E. Y. was funded by MEXT Grant-in-Aid for the ``Program for Leading Graduate Schools."

\section*{References}

%\bibliographystyle{iopart-num}  

%\bibliography{attm}

\providecommand{\newblock}{}

\appendix

\section{$n$th moment of $Z(t)$}
The $n$th derivative of $\hat{P}(k,s)$ satisfies the following recursion relation:
%\begin{widetext}
\begin{eqnarray}
\hat{P}_D^{(n)}(k,s) = \frac{1}{1-\hat{\psi}_D(k,s)} \left[ \sum_{i=1}^{n-1} c_{n,i}\hat{P}_D^{(i)}(k,s) \hat{\psi}_D^{(n-i)}(k,s) + 
\hat{P}_D(k,s) \hat{\psi}_D^{(n)}(k,s) \right. \nonumber\\ 
  \left. + \int_0^\infty d\tau \rho(\tau) \hat{\Psi}_D^{(n)}(k,s;\tau)
\right],
%\tag{S10}
\end{eqnarray}
%\end{widetext}
where $c_{n,i}=c_{n-1,i}+c_{n-1,i-1}$ ($i=2, \ldots, n-2$) and $c_{n,n-1}=c_{n,1}=n$.
\if0 
\begin{align}
\int_0^\infty d\tau \rho(\tau) \hat{\Psi}_D^{(n)}(0,s;\tau) &=& {\color{red}(-1)^n}\int_0^\infty d\tau \rho(\tau) D_\tau^n \int_0^\tau t^n e^{-st}dt \nonumber\\
&= {\color{red}(-1)^n}\int_0^\infty dt \int_t^\infty d\tau \rho(\tau) D_\tau^n t^n e^{-st} \nonumber\\
&\sim {\color{red}(-1)^n}c \int_0^\infty dt e^{-st} t^n \int_t^\infty \tau^{-1 -\alpha +n (\sigma-1)} d\tau \nonumber\\
&\sim \frac{{\color{red}(-1)^n}c}{(n(1-\sigma) + \alpha)\Gamma(1-\alpha + n\sigma)}\frac{1}{s^{1-\alpha + n\sigma}}.
%\tag{S11}
\end{align}
\fi
Here, we assume that
\begin{equation}
\hat{P}_D^{(i)}(0,s) \sim (-1)^n\frac{M_i(\alpha,\sigma)}{|\Gamma(-\alpha)|} \frac{1}{s^{1+i \sigma}}.
%\tag{S11}
\end{equation}
It follows that 
\begin{equation}
\fl
\hat{P}^{(n)}(0,s) %&=  \left[\sum_{i=1}^{n-1} c_{n,i} (-1)^{\color{red}n}\frac{M_i(\alpha,\sigma)}{|\Gamma(-\alpha)|}  
%\frac{\Gamma((n-i)\sigma - \alpha)}{|\Gamma(-\alpha)|} + \frac{\Gamma(n\sigma - \alpha)}{|\Gamma(-\alpha)|}
%+  \frac{\Gamma(n\sigma - \alpha +1)}{|\Gamma(-\alpha)|(n+\alpha-n\sigma)} \right] \frac{1}{s^{1 + n\sigma}} \nonumber\\
= \left[\sum_{i=1}^{n-1} c_{n,i} (-1)^{n}M_i(\alpha,\sigma) \frac{\Gamma((n-i)\sigma - \alpha)}{|\Gamma(-\alpha)|} 
+  \frac{n}{n+\alpha-n\sigma} \Gamma(n\sigma - \alpha) \right] 
\frac{1}{|\Gamma(-\alpha)| s^{1 + n\sigma}}. 
%\tag{S12}
\end{equation}
Therefore, 
\begin{equation}
M_n (\alpha,\sigma) = (-1)^{n}\sum_{i=1}^{n-1} c_{n,i} M_i(\alpha,\sigma) \frac{\Gamma((n-i)\sigma - \alpha)}{|\Gamma(-\alpha)|} 
+ \frac{n}{n+\alpha-n\sigma}\Gamma(n\sigma - \alpha).
%\tag{S13}
%\label{Mn}
\end{equation}

\end{document}